# Carrier duality from the convergence of Dirac fermions and high-order van Hove singularities

Meng Lyu[1†], Kaiyi Zhai[2†], Nikolai Peshcherenko[3], Junyan Liu[1], Jinying Yang[1], Subir Sen[1], Binbin Wang[1], Langsheng Ling[4], Zhaosheng Wang[4], Gang Li[1], Jieyi Liu[5,6], Yang Xu[1], Xiyang Li[1], Claudia Felser[3], Yang Zhang[3], Wujun Shi[7,8*], Lexian Yang[2,9*], Enke Liu[1*]

[1]Beijing National Laboratory for Condensed Matter Physics, Institute of Physics, Chinese Academy of Sciences, Beijing 100190, China.

[2]State Key Laboratory of Low Dimensional Quantum Physics, Department of Physics, Tsinghua University, Beijing 100084, China.

[3]Max Planck Institute for Chemical Physics of Solids, Dresden 01187, Germany.

[4]Anhui Province Key Laboratory of Condensed Matter Physics at Extreme Conditions, High Magnetic Field Laboratory, Chinese Academy of Sciences, Hefei 230031, China.

[5]Diamond Light Source, Harwell Science and Innovation Campus, Didcot OX11 0DE, UK.

[6]Department of Physics, Clarendon Laboratory, University of Oxford, Parks Road, Oxford OX1 3PU, UK.

[7]State Key Laboratory of Quantum Functional Materials, Center for Transformative Science, ShanghaiTech University, Shanghai 201210, China.

[8]Shanghai High Repetition Rate XFEL and Extreme Light Facility (SHINE), ShanghaiTech University, Shanghai 201210, China.

[9]Frontier Science Center for Quantum Information, Beijing 100084, China.

†These authors contributed equally to this work.

*Corresponding author. Emails:

shiwujun@shanghaitech.edu.cn; lxyang@tsinghua.edu.cn; ekliu@iphy.ac.cn

**The convergence of highly mobile Dirac fermions and flat-band heavy electrons offers a paradigm for emergent quantum phenomena. However, experimental realization of such intriguing state remains elusive. In this study, we report a dual carrier regime in the kagome metal $Co_3In_2S_2$, wherein the charge transport is governed by high-mobility electrons while the thermodynamic responses exhibit heavy-electron behavior. This exotic duality arises from the coexistence of topological Dirac fermions and flat-band high-order ($4^{th}$-order) Van Hove singularities (HOVHSs) at the Fermi surface, as revealed by magnetic-torque quantum oscillation and angle-resolved photoemission spectroscopy. The interaction between Dirac and HOVHS-derived carriers can be captured by a minimal two-pocket model, manifesting as sublinear resistivity above ~100 K and field-induced non-Fermi-liquid behavior at low temperatures. Our study establishes $Co_3In_2S_2$ as a model platform for exploring many-body physics at the intersection of topology and correlation effects, and provides a foundational framework for designing quantum materials hosting diverse emergent states.**

# I. INTRODUCTION

The strong interplay between mobile conduction carriers and localized electrons can engender a rich spectrum of intriguing emergent phenomena, such as high-temperature superconductivity [1,2], quantum criticality [3,4], and strongly correlated topological quantum phases [5,6]. Among these, flat band systems serve as paradigmatic platforms where kinetic energy suppression enables the realization of exotic correlated states. Two canonical classes have been extensively studied: (1) heavy-fermion systems, where localized *f*-electrons hybridize with conduction electrons, giving rise to a renormalized flat band near the Fermi level ($E_F$) (Fig. 1(a)) and forming heavy quasiparticles that dominate the low-temperature physical properties [7]; and (2) frustration-lattice materials, where destructive quantum interference effects quench kinetic energy to induce band flattening. In some moiré systems, the coexistence of Dirac bands and flat bands at $E_F$ fosters interband interactions between light Dirac fermions and heavy quasiparticles (Fig. 1(b)), which have been linked to superconductivity stabilization in magic-angle twisted bilayer graphene [8-9]. These experimental achievements prompt a natural question: can the light-heavy interplay be extended to other flat-band platforms, and to what extent can the two types of carriers exhibit their inherent properties?

This puzzle motivates our investigation of correlated materials featuring van Hove singularities (VHSs), which give rise to locally flat bands, thus providing a versatile platform to explore the interplay between heavy localized electrons and light conduction electrons [10]. In particular, the effective mass can be further enhanced by the high-order VHS (HOVHS) [11]. Unlike ordinary VHSs manifested as first-order derivative discontinuities and logarithmic divergence of density of state (DOS), HOVHS is a nontrivial singularity caused by higher-order derivative discontinuities, leading to a power-law divergence in the DOS. In theory, HOVHSs can induce many-body physics and intriguing emergent phenomena, such as strong electron correlation effects [11,12], superconducting pair-density-wave and supermetals [13,14]. Nevertheless, the experimental investigation of HOVHS remains inadequate [15-18], calling for other materials with HOVHS, especially those exhibiting flat-band effect (Fig. 1(c)).

Recent studies have demonstrated that kagome materials, characterized by their unique electronic structure of flat bands, VHSs, and Dirac points, serve as a fertile platform for realizing emergent quantum states [17-24]. Among them, the magnetic Weyl semimetal $Co_3Sn_2S_2$ exhibits particularly intriguing properties, such as giant anomalous Hall effects [25]. In this work, we investigate the thermodynamic, transport, and quantum oscillation, as well as the electronic structure, of a kagome metal $Co_3In_2S_2$, which is isostructural with $Co_3Sn_2S_2$ (rhombohedral structure with space group R-3m) [25]. We reveal an exclusive carrier-duality behavior of high mobility in transport and heavy electrons in thermodynamics at low temperatures. Furthermore, the distinct electronic structure from the convergence between HOVHS and Dirac fermions at $E_F$ was directly visualized by high-resolution angle-resolved photoemission spectroscopy (ARPES), which paves the way for the novel properties of this compound.

## II. RESULTS AND DISCUSSION

### A. Heavy electron, sublinear resistivity, and non-Fermi liquid behavior

The structure of $Co_3In_2S_2$ consists of $Co_3In$ kagome layers that are vertically sandwiched by two Sulfur atomic planes and one indium atomic plane (Fig. 2(a)).

$Co_3In_2S_2$ exhibits paramagnetic behavior down to 2 K, which is distinct from the ferromagnetic ground state of $Co_3Sn_2S_2$ [25]. Remarkably, it demonstrates an unexpected high Sommerfeld coefficient $\gamma$ = 42 mJ mol$^{-1}$ K$^{-2}$ (Fig. 2(b)). This is significantly larger than that of $Co_3Sn_2S_2$ but comparable to canonical flat-band systems $Ni_3In$ [22] and $CuV_2S_4$ [26], thereby suggesting a heavy-electron-like behavior. The magnetic susceptibility curves (see Supplementary Material [27]) follow a Curie-Weiss temperature dependence with a pronounced Pauli susceptibility component ($\chi_{\mathrm{P}}$), enabling the calculation of the Wilson ratio via $R_{\mathrm{W}} = \left(\frac{\pi^2\kappa_{\mathrm{B}}^2}{3\mu_{\mathrm{B}}^2}\right)\left(\frac{\chi_{\mathrm{P}}}{\gamma}\right) = 2.1$. This value aligns closely with the characteristic range of heavy-fermion materials [27, 28], further corroborating the presence of heavy electrons in $Co_3In_2S_2$. It is noteworthy that while $Co_3In_2S_2$ shows no evidence of magnetic ordering, a slight Fe doping induces an antiferromagnetic transition in $(Co_{1-x}Fe_x)_3In_2S_2$ [29] . The previous study also suggested of a fragile antiferromagnetic order in the pristine $Co_3In_2S_2$ with different growth method [30]. Taken together, these findings establish $Co_3In_2S_2$ as a correlated material proximate to the magnetic ordering.

Figure 2(c) presents the zero-field temperature-dependence of the in-plane resistivity ($\rho_{\mathrm{xx}}$), exhibiting a pronounced sublinear temperature dependence at high temperatures, following the scaling of $\rho_{\mathrm{xx}} \propto T^{0.53}$. At low temperatures, the resistivity follows $\rho_{\mathrm{xx}} \propto T^{1.93}$, closely approaching the canonical Fermi liquid (FL) behavior. However, the application of a magnetic field ($B \| \mathrm{c}$) systematically disrupts this FL state. High-resolution $\rho_{\mathrm{xx}}(B)$ measurements (inset, Fig. 2(d)) reveal a progressive emergence of positive magnetoresistance with increasing field strength. Most notably, at $B$ = 16 T, the resistivity deviates significantly from FL predictions, obeying the scaling of $\rho_{\mathrm{xx}} \propto T^{1.5}$ (Fig. 2(d)), a hallmark signature of non-Fermi liquid (NFL) behavior. To quantitatively characterize this crossover, we fit $\rho_{\mathrm{xx}}$ to the phenomenological form$\rho_{\mathrm{xx}} = \rho_0 + AT^n$, where $\rho_0$ denotes the residual resistivity, yielding the exponent $n$ as a function of magnetic field, as shown in Fig. 2(e). The plot demonstrates a systematic reduction of $n$ with increasing field, stabilizing at a critical value $n \approx 1.5$ in the high-field regime, a signature of NFL scaling. This exponent is further cross-validated through separate analysis of the inset curves $\ln(\rho_{\mathrm{xx}} - \rho_0)$ vs $\ln T$, where the slope $n = d\ln(\rho_{\mathrm{xx}} - \rho_0)/d\ln T$ directly maps the temperature dependence (Fig. 2(f)). The resulting $n(T, H)$ plot highlights a distinct regime with $n \approx 1.5$, unambiguously confirming the NFL behavior. This scaling exponent can be compared with recent reports of strong-correlation-induced NFL resistivity in the pyrochlore lattice $CuV_2S_4$ ($n \approx 1.6$) [26] and the kagome metal $Ni_3In$ ($n \approx 1$ under low field or moderate pressure) [22], suggesting that $Co_3In_2S_2$ may host analogous flat-band-driven NFL physics.

**B. Coexistence of Dirac and HOVHS fermions on the Fermi surface**

In order to comprehend the above novel properties, we performed high-resolution ARPES experiment to investigate the electronic structure of $Co_3In_2S_2$ (Fig. 3). The Fermi surface exhibits a six-fold rotation symmetry as expected for a kagome lattice (Fig. 3(a)). We observe a warped hexagonal electron pocket around the Brillouin zone center, together with broad spectral weight distribution near the $\bar{M}$ point. Since In atom has one *p*-electron less than Sn atom, replacing Sn by In lowers $E_F$ of $Co_3In_2S_2$ compared to that of $Co_3Sn_2S_2$, resulting in distinct Fermi surface structures of the two sister compounds [20].

Figures 3(b) and 3(c) compare the measured band dispersions along high-symmetry directions with the surface-projected calculations using density-functional theory (DFT), showing an overall agreement. Noticeably, we observe two Dirac points near $E_F$ along $\bar{\Gamma}\bar{M}$ and $\bar{\Gamma}\bar{K}$ (red circles in Fig. 3(c)), which are better visualized in the zoom-in plot in Fig. 3(d). These quickly dispersing Dirac fermions close to $E_F$ (~ -18 meV and -30 meV along $\bar{\Gamma}\bar{M}$ and $\bar{\Gamma}\bar{K}$ respectively) will significantly contribute to the transport properties of the system.

In addition to the Dirac fermions, we observe sharp flat band near the $\bar{M}$ point (Fig. 3(b)). Figure 3(e) shows the band dispersions measured along four momentum directions indicated in Fig. 3(a). Exceptionally sharp flat bands appear precisely at $E_F$ in cuts #3-#6 (Fig. 3(e)). These bands, characterized by their remarkably narrow width and attributed to *d*-orbital electrons, are perfectly poised to serve as an ideal platform for studying strongly correlation physics [22,26]. Figure 3(f) further presents a detailed examination of the band dispersion near the $\bar{M}$ point. To visualize the spectral features above $E_F$, the data along $\bar{M}\bar{\Gamma}$ were divided by Fermi-Dirac distribution function. It is noteworthy that the dispersions along $\bar{M}\bar{K}$ and $\bar{M}\bar{\Gamma}$ are hole-like and electron-like respectively, converging at a VHS at $\bar{M}$, in good agreement with the theoretical calculations in Supplementary Material [27].

To accurately extract the band dispersion near the VHS, we fit the energy distribution curves (EDCs) to Lorentzian functions to extract the flat band dispersion. Interestingly, the band dispersion is well-described by a quartic function ($E \propto k^4$, blue solid line in Fig. 3(f)), while a quadratic function ($E \propto k^2$, dashed red line in Fig. 3(f)) clearly deviates the experimental data, identifying a HOVHS at $\bar{M}$ point (see Supplementary Material [27]). A polynomial fit of $E(k) = ak^2 + bk^4$ yields $a = -10.6 \pm 19.9$ meV Å$^2$ and $b = 5594 \pm 410$ meV Å$^4$. The vanishing ratio $a/b \approx -0.0019$ Å$^{-2}$ suggests that the quadratic coefficient is consistent with zero within the experimental uncertainty. Moreover, the crossover energy below which the quadratic term would dominate can be estimated by $E_c = 2a^2/b \approx 0.4$ meV, which falls far below the energy scales relevant to the electronic transport. Consequently, the quadratic contribution is negligible across the experimentally relevant energy and momentum window.

Figure 3(g) further demonstrates the DOS of the HOVHS evaluated from the reconstructed band dispersions along two orthogonal momentum directions crossing $\bar{M}$. The calculated DOS is well fitted by $D(E) \propto (E - E_{\mathrm{VHS}})^{-\beta}$, with $\beta = 0.68 \pm 0.02$. This power-law DOS divergence stems from HOVHS states extending over a broad momentum region near the $\bar{M}$ point at the $E_F$ (Figs. 3(e) and 3(f)), which may give rise to a heavy-electron behavior, consistent with the enhanced Sommerfeld coefficient (Fig. 2(b)).

**C. Heavy-electron feature of the HOVHS from quantum oscillation**

To further probe the flat-band nature of the HOVHS, we turn to quantum oscillation measurements. High-field magnetoresistance (up to 30 T) and magnetization (up to 35 T) at 2 K show no discernible quantum oscillation features (see Supplementary Material [27]), indicating that accessing the quantum oscillation regime may require much lower temperatures. We therefore performed magnetic torque measurements in a dilution refrigerator down to 40 mK and up to 29 T. Figure 4 displays the quantum oscillation properties of $Co_3In_2S_2$ with the magnetic fields perpendicular to the kagome plane. Clear quantum oscillations emerge above 12 T at the base temperature and are rapidly suppressed above 1 K (Fig. 4(a)). The fast Fourier transform (FFT) of the oscillatory component (Fig. 4(b)) reveals several distinct frequencies at approximately 2248 T, 4604 T, and 4996 T, labeled as $\alpha_1$ to $\alpha_3$, which correspond to extremal cross-sectional areas of the Fermi surface. Fitting the temperature dependence of each FFT amplitude to the Lifshitz–Kosevich formula (Fig. 4(c)), we extract cyclotron effective masses up to $m^* = 9.8\ m_0$. This value is comparable to those found in typical *f*-electron heavy-fermion systems [31], providing direct evidence for heavy-electron quasiparticles in $Co_3In_2S_2$. This conclusion is further supported by independent examinations of the HOVHS band mass at $E_F$ from our ARPES measurements and DFT calculations (see Supplementary Material [27]), which also yield an enhanced effective mass of $m^* \approx$ 6-9 $m_0$, as summarized in Fig. 4d. Combined with the large Sommerfeld coefficient obtained from specific-heat measurements, the convergence of these independent probes confirms that the HOVHSs dominate the thermodynamic response and serve as flat bands that give rise to the heavy-electron behavior.

**D. Dual nature of Dirac and HOVHS fermions.**

Building upon the established electronic band structure of $Co_3In_2S_2$, in which both Dirac points and HOVHSs are located at $E_F$ (Fig. 5(a)) and depicted in the entire first Brillouin zone (Fig. 5(b)), we employ the minimal semiclassical two-pocket (Dirac cone + HOVHS) model [32] to elucidate the anomalous transport observed in this system. Within this model, the weakly dispersing HOVHS states act as a momentum-relaxing reservoir for the high-mobility Dirac electrons. In the experimentally relevant regime where the HOVHS lies sufficiently close to $E_F$ ($T \gg \mu_V$, where $\mu_V$ is the HOVHS offset from $E_F$), the momentum-space size of the thermally activated electron cloud near the HOVHS grows with temperature as $q_T \propto (T/\mu_V)^{1/\alpha}$, where $\alpha > 2$ characterizes the weak dispersion of the HOVHS band. Internode electron–electron (e-e) scattering between Dirac and HOVHS carriers then yields a sublinear resistivity scaling, since the resulting scattering rate scales as $\tau_e^{-1} \propto T^{3/\alpha}$. At temperatures above approximately 100 K, the measured resistivity follows $\rho_{xx} \propto T^{0.53}$ (Fig. 2(c)). This behavior cannot be attributed to conventional electron–phonon scattering, which typically produces $n \geq 1$, nor to impurity scattering, which contributes only a temperature-independent term. Within the two-pocket framework, the $n$ exponent can be predicted from the DOS divergence parameter $\beta$ (Fig. 3(g)), yielding $n = 3/2(1-\beta) = 0.48$. This value is in reasonable agreement with the experimentally fitted exponent of 0.53. In addition, the emergence of NFL behavior could also be understood as a consequence of the enhanced interplay between HOVHS and Dirac electrons. In the presence of magnetic field, Zeeman splitting drives one HOVHS branch towards $E_F$, making the sublinear behavior more pronounced as an NFL feature (see Supplementary Material [27]). While the minimal two-pocket model successfully captures the field-induced suppression of the resistivity exponent from $n = 2$ to $n \approx 1.5$ via Dirac–HOVHS inter-pocket scattering at low-to-intermediate fields, the persistence of this exponent at higher fields likely reflects field-induced modifications of the inter-pocket coupling and correlation effects beyond the present semiclassical framework.

In addition, it is noticed that the residual resistivity ($\rho_0$) of $Co_3In_2S_2$ remains exceptionally low even within the NFL state, with values spanning $\rho_0 \sim 0.2 - 0.5$ μΩ cm (Fig. 2(d)), 2–3 orders of magnitude smaller than those of NFL frustrated metal of $Ni_3In$ [22] and $CuV_2S_4$ [26]. This ultralow $\rho_0$ is comparable to that of the ferromagnetic metal MnSi [33], in which a similar NFL state with $\rho_{xx} \propto T^{1.5}$ under pressure was observed. Such a small $\rho_0$ could imply a high carrier mobility ($\mu$) due to the presence of the topological bands [34, 35]. To quantify this, the two-band model [36] is employed to jointly fit the Hall and longitudinal conductivity data (see details in Supplementary material [27]). At $T$ = 2 K, the best fit yields two distinct carrier populations: a high-mobility pocket with $\mu_1 = 8.4 \times 10^3$ cm$^2$ V$^{-1}$ s$^{-1}$ and a slightly lower-mobility pocket with $\mu_2 = 1.03 \times 10^3$ cm$^2$ V$^{-1}$ s$^{-1}$. Given the heavy-electron character inferred from thermodynamic and quantum oscillation measurements (e.g., large $m^*$), the coexistence of high mobility suggests a dominant role of Dirac fermions in electrical transport, while the sublinear resistivity reveals the concurrent participation of strongly correlated electrons. Further, we calculate the Kadowaki–Woods ratio via $R_{\mathrm{KW}} = A/\gamma^2$ (see Supplementary Material [27]), which yields a $R_{\mathrm{KW}} \approx 1.4 \times 10^{-6}$ μΩ cm mol$^2$ K$^2$ mJ$^{-2}$. This value falls between those typical of conventional transition metals and canonical heavy-fermion systems. As is the case for the van der Waals heavy-fermion compound CeSiI [37], the relatively small $A$ coefficient, combined with the large $\gamma$, directly reflects the coexistence of light, conduction carriers that dominate transport and heavy, renormalized flat-band quasiparticles that govern the thermodynamics. These observations establish $Co_3In_2S_2$ as a unique system exhibiting dual electronic states: high-mobility light Dirac fermions and heavy HOVHS quasiparticles, mediated by the correlation between HOVHSs and Dirac points.

The interplay between flat-band HOVHSs and Dirac fermions in $Co_3In_2S_2$ evokes analogies with strong e–e correlations in heavy-fermion compounds. In such system, localized *f*-electrons interact freely with itinerant *s*-, *p*- or *d*-derived electrons above the coherent temperature, and are subsequently screened by conduction electrons at lower temperatures, to form heavy quasiparticles via *c*-*f* hybridization within a Kondo lattice. By stark contrast, $Co_3In_2S_2$ features a kagome-lattice-driven interaction mechanism in which the electronic landscape is dominated by Co-3*d* states at the Fermi level (Fig. 5(c)). Here, localized Co-3*d* electrons derived from HOVHSs immersed in the Fermi sea engage in interacting with fast 3*d* electrons originating from Dirac points (Fig. 5(c) inset). This interaction processes are demonstrated to govern the electronic transport, exhibiting distinct temperature-field-dependent behaviors that reflect the evolution of many-body correlations, as illustrated in the phase diagram of Fig. 5(d). Three typical regions are identified. At low temperatures and fields, a dual state of heavy electrons and light Dirac fermions emerges, and the interplay between the two is relatively weak, with transport being dominated by the Dirac fermions and exhibiting a FL feature. In conditions of elevated temperature, the HOVHS-Dirac interaction becomes predominant, thus resulting in sublinear metallic behavior. This sublinear behavior at low temperatures evolves to a typical NFL phenomenon with increasing magnetic fields. The three different predominant interactions in the phase diagram can also be clearly demonstrated by Koler's scaling rule of magnetoresistance [38], as shown in the (see Supplementary Material [27]). Notably, a *d*-electron kagome Kondo lattice has recently been realized in $CsCr_6Sb_6$ [39], exhibiting an extremely flat band in ARPES and hallmark Kondo resistivity upturn. Yet the correlated state in $Co_3In_2S_2$, born from the von Hove renormalized Dirac Fermions, diverges fundamentally from conventional Kondo physics: no Kondo effect is observed. It is emphasized that the kagome geometry, when combined with HOVHS-related band flattening, enables a novel coupling paradigm that transcends the canonical Kondo lattice picture. Consequently, $Co_3In_2S_2$ appears to host a previously unreported quantum ground state, where the synergy of Dirac fermions and HOVHS-derived heavy electrons yields anomalous transport without invoking localized magnetic moments or Kondo screening.

To further demonstrate the distinctiveness of $Co_3In_2S_2$, Fig. 5(e) presents the Sommerfeld coefficient dependence of the mobility across diverse kagome materials [22, 23, 25, 40-50]. This plot reveals a general trend: $\mu$ decreases with increasing $\gamma$, which stems from the fundamental relationship between $\gamma$ (a measure of Fermi-liquid correlation strength, proportional to effective mass $m^*$) and transport properties (heavy electrons move sluggishly). Consequently, the transport dominance of kagome materials can be naively classified by their position in the $\mu$-$\gamma$ plot. The top-left corner (e.g., CoSn) is dominated by Dirac fermions (high $\mu$, low $\gamma$), while the bottom-right corner (e.g., $CsCr_3Sb_5$) is dominated by flat-band-derived heavy electrons (low $\mu$, high $\gamma$). Remarkably, $Co_3In_2S_2$ stands out as an exclusive material, exhibiting novel duality of both high mobility and large effective mass, owing to its distinctive electronic structure at $E_F$ (Fig. 5(a)). Finally, we note that the mobility is also inherently sensitive to sample quality, and higher-quality $Co_3In_2S_2$ samples can even achieve a mobility exceeding 10000 cm$^2$ V$^{-1}$ s$^{-1}$ in the early report [30].

## III. CONCLUSION

In conclusion, we identify the compound $Co_3In_2S_2$ as an exceptional kagome system by unifying the correlated HOVHSs and topological Dirac fermions on the Fermi surface, which gives rise to a striking carrier duality of heavy-electron quasiparticles and high-mobility carriers. The sublinear transport behavior and the NFL behavior further establish $Co_3In_2S_2$ as a paradigmatic system to explore many-body physics and correlated topology. Our findings pave the way for manipulating quantum states via band engineering, with implications for thermodynamic behaviors and quantum effects in novel materials, such as efficient heat transmission in advanced thermoelectrics [51, 52], or strong correlation in unconventional superconductivity [8, 9, 23].

# IV. METHODS

### A. Crystal growth and characterizations

$Co_3In_2S_2$ single crystals were grown by the high-temperature In-flux method. High purity chunks of cobalt, indium, and sulfur were loaded into an alumina crucible and further sealed in the quartz tube under high vacuum. Typical large hexagonal single crystalline $Co_3In_2S_2$ platelets with dimensions of 2~10 mm were obtained (see details in Supplementary Material [27]).

The chemical composition of the as-grown crystal was determined by energy dispersive X-ray spectroscopy (EDS) using a Hitachi S-4800 scanning electron microscope (SEM), with a ratio of Co: In: S ~ 3: 2: 2. The single crystal orientation was determined by X-ray diffraction (XRD) on a Bruker D8 Venture diffractometer at room temperature using Cu-$K_\alpha$ radiation ($\lambda$ = 0.71073Å). The only presence of Bragg peaks of (00$l$) indicates that the large, exposed surface of the obtained crystal (Supplementary Material [27]) is the ab plane.

### B. Specific heat measurements

The specific heat was measured over the range of 2-200 K using a Quantum Design Physical Property Measurement System (PPMS) with a standard relaxation method. The addenda data were measured in advance with N grease. Subsequently, single crystals of $Co_3Sn_2S_2$ (10.8 mg) and $Co_3In_2S_2$ (12 mg) were adhered to the heat capacity puck using these N grease for the measurement. The measurements were carried out for three times at each temperature, and the final $C(T)$ data were obtained by averaging (see Supplementary Material [27]). The temperature rise was set to the minimal value of 1% in the PPMS to ensure optimal temperature resolution.

### C. Magnetization measurements

DC Magnetic susceptibility and magnetization measurements were performed in a Magnetic Property Measurement System (MPMS) using a Superconducting Quantum Interference Device (SQUID) magnetometer. The measurements were performed with the magnetic field applied parallel and perpendicular to the c-axis, using a copper sample holder. The magnetic susceptibility $\chi(T)$ were fitted by the formula $\chi(T) = \chi_0 + C/(T-\theta)$ (see Supplementary Material [27]). The fitting results are all presented in the inset of the figures.

**D. Electrical transport measurements**

The electrical transport measurements were conducted in the PPMS at temperatures ranging from 2 to 300 K, with current in the ab plane and magnetic fields oriented parallel to the c-axis ($I\parallel$a and $B\parallel$c). The standard six-probe method was applied for the longitudinal resistivity and the Hall effect measurements. In order to eliminate the influence of the misalignment of the lead contacts, all the magnetoresistance and Hall effect measurements were conducted by scanning both negative and positive magnetic fields. In order to guarantee the precision of the data, three samples (designated S1, S2, and S3) were measured using the same sequence. The results of S1 were presented in the main text, with S2 and S3 presented in the Supplementary Material [27]. The measured samples were cut and polished into a regular bar shape with dimensions of approximately 2×0.5×0.1 $mm^3$. The 16-T resistivity measurements were carried out at the F2 station of the Synergetic Extreme Condition User Facility (SECUF), Beijing, China.

**E. High-field magnetization, magnetoresistance, and torque measurements**

The high-field magnetization (Supplementary Material [27]) at 2 K was conducted in a water-cooled magnet was conducted in a water-cooled magnet with a maximum field up to 35 T using a hybrid magnet at the High Magnetic Field Laboratory of the Chinese Academy of Sciences, Hefei, China. The high-field longitudinal magnetoresistance at 2 K and magnetic torque were measured in static magnetic fields of up to 29 T in a superconductor magnet at the A3 station of SECUF (see Supplementary Material [27]). The magnetoresistance was conducted using a standard four-probe method in a $^4$He cryostat with $I\parallel$a and $B\parallel$c. The current was set at 3 mA, modulated at a frequency of 27.7 Hz using a Keithley 6221. The voltage was measured using a SR830 Lock-In Amplifier. The magnetic torque was measured at dilution refrigerator down to 40 mK.

**F. Angle-resolved photoemission spectroscopy (ARPES)**

High-resolution ARPES measurements were conducted at beamline BLOCH in MAX IV (proposal No. 20230668). The samples were cleaved in-situ under ultra-high vacuum below $8\times10^{-11}$ mbar. Data were collected with a Scienta DA30L (DA30) electron analyzer. The total energy and angular resolutions were set to 15 meV and 0.2°, respectively. Data in Fig. 3(b), (f), and (g) were collected using 70 eV photons and data in Fig. 3(e) were collected using 48 eV photons.

### G. DFT calculations

The first-principles calculations were performed using the Vienna ab initio Simulation Package (VASP) [53]. The interactions between the valence electrons and ion cores are described by the projector augmented wave method [54, 55], and exchange-correlation potential is formulated by the generalized gradient approximation with the Perdew-Burke-Ernzerhof (PBE) scheme [56]. The Γ-centered 16×16×16 k-points are used for the first Brillouin-zone sampling. The spin-orbit coupling (SOC) is included in all the calculations. The tight-binding Hamiltonian was constructed using the maximally localized Wannier functions which was provided by Wannier90 packagen [57]. The surface states were calculated by the surface Green's function method [58] based the tight-binding Hamiltonian. The relaxed and experimental lattice constant (Inorganic Crystal Structure Database no. 5436) were used in the calculations.

### H. Model analysis of the resistivity

What follows is a description of the theoretical model that we use to understand the high-temperature sublinear scaling of resistivity, as well as the magnetic field-dependent NFL resistivity behavior observed at low temperatures. These observed phenomena are attributed to the inter-pocket e-e scattering channel description for Kagome metals [32]. Namely, band structure of Kagome metals is known to host a higher-order VHS with dispersion relation $E_{\mathrm{VHS}}(q) \propto q^{\alpha}, \alpha > 2$ positioned close to the Fermi level. In that case, the VHS electrons do not provide direct contribution to the charge current due to low group velocity. However, its large density of states enhances the scattering of fast current-carrying electronic states belonging to other pockets at the VHS pocket electrons. The transmitted momenta are nevertheless limited by the size of thermally activated electronic cloud $q_{act}$ near the VHS point ($E_{\mathrm{VHS}}(q_{act}) \propto k_{\mathrm{B}}T$), since only the electrons close to the VHS would contribute to this scattering channel. The estimated sublinear scaling for $q_{act}$ then provides $q_{act} \propto T^{1/\alpha}$, allowing for a sublinear behavior of scattered electrons' phase volume with temperature. This, in turn, could be shown to result in sublinear scaling of the rate of electron-electron scattering. The resulting sublinearity exponent would depend on the details of band dispersion near VHS.

**ACKNOWLEDGMENTS**

This work was supported by the State Key Development Program for Basic Research of China (No. 2022YFA1403400 and 2022YFA1403100), the National Natural Science Foundation of China (Nos. 12595332, 12604256, and 12274251), the Synergetic Extreme Condition User Facility (SECUF), the Scientific Instrument Developing Project of CAS (No. ZDKYYQ20210003), and China-Germany Joint Research Center on Quantum Materials and Physics under Extreme Conditions. We thank the WM5 of the Steady High Magnetic Field Facility, CAS (https://cstr.cn/31125.02.SHMFF.WM5) for the assistance on the experiment.

E.L. and L.Y. conceived and supervised the project. M.L. grew the single crystals and performed the structural and chemical characterizations. K.Z. conducted the ARPES measurements under guide of L.Y., W.S., and J.L. performed the first-principles calculations. N.P and Y.Z. performed the modelling. M.L., J.Y., S.S., B.W., Y. X., and X.L. performed magnetic, thermodynamic, and transport measurements. M.L., L.L., Z.W., performed the high-field magnetization measurements. M.L., and G.L. carried out the high-field magnetic torque and magnetoresistance measurements. E.L., L.Y., W.S., Y.Z., C.F., M.L., and K.Z. analyzed and discussed the experimental data. M.L., K.Z., L.Y., and E.L. wrote this manuscript with input from all authors.

**DATA AVAILABILITY**

The data are available from the authors upon reasonable request.

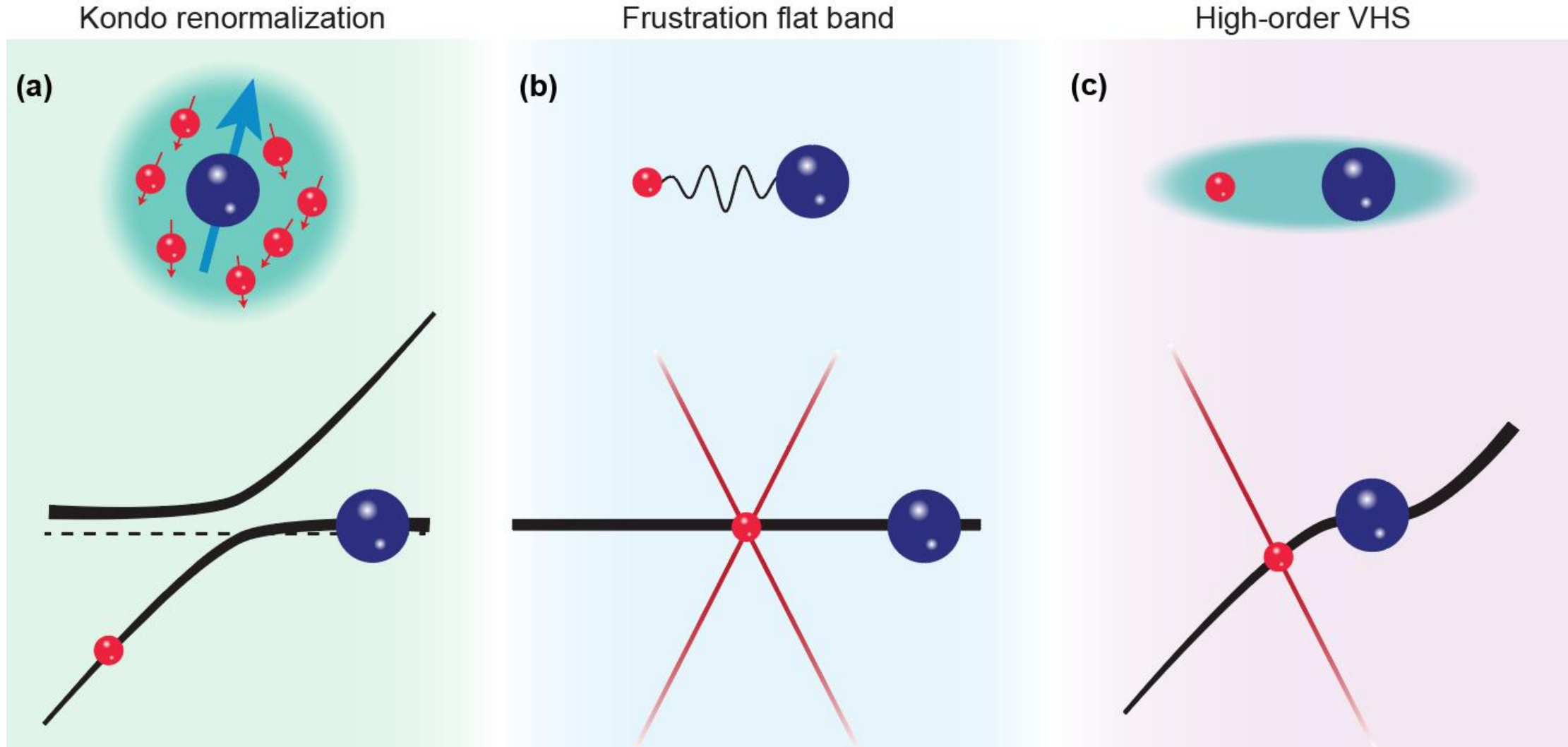

**Fig. 1 | Landscape of the flat-band systems with entangled light and heavy electrons.** **(a)** Hybridization between conduction carriers and localized *f* electrons in a Kondo lattice, resulting in renormalized flat bands. (**b)** Interaction between a Dirac band and the hopping frustration-induced flat band, as exemplified in the frustrated lattices or moiré superlattice materials. (**c)** High-order van Hove singularity (VHS), representing a novel class of flat bands and driving novel correlated electronic states through their interplay with Dirac fermions.

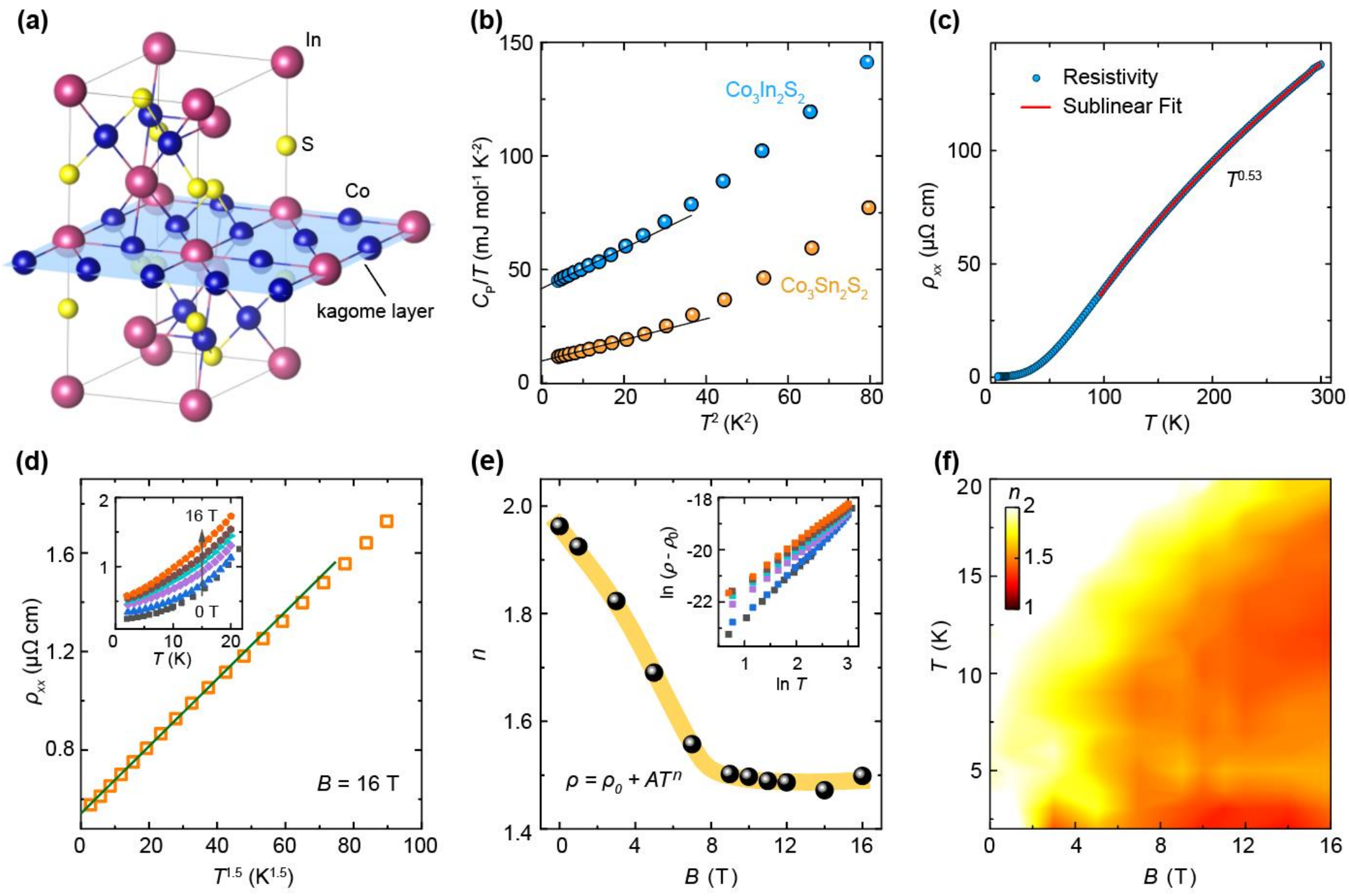

**Fig. 2 | Novel thermodynamic and transport properties of $Co_3In_2S_2$. (a)** Schematic illustration of the crystal structure of $Co_3In_2S_2$. The blue plane highlights the $Co_3In$ kagome layer. (**b**) The largely enhanced Sommerfeld coefficient ($\gamma$) of $Co_3In_2S_2$ with $\gamma$ = 42 mJ mol$^{-1}$ K$^{-2}$, in comparison to that of $Co_3Sn_2S_2$ with $\gamma$ = 11 mJ mol$^{-1}$ K$^{-2}$. The black line is a linear fit to the data at low temperatures. (**c**) Sublinear resistivity behavior at temperatures above 100 K with $\rho_{\mathrm{xx}} \propto T^{0.53}$. At low temperatures, a canonical Fermi liquid behavior was observed, with $\rho_{\mathrm{xx}} \propto T^{1.93}$. (**d**) NFL behavior with the resistivity in a $T^{1.5}$ scale under a magnetic field of 16 T along the c axis. The green line is the linear fit to the data, confirming the $T^{1.5}$ dependence of the resistivity below 20 K. The inset shows the raw data of temperature-dependent resistivity in selected magnetic fields of $B$ = 0, 3, 7, 10, 12, 16 T. (**e**) Magnetic field dependence of the exponent $n$, which is extracted by fitting the data in the inset of (**d**) to the formula of $\rho = \rho_0 + AT^n$. The inset shows the $\ln(\rho_{\mathrm{xx}} - \rho_0)$ vs $\ln T$ plot of the resistivity curves under different magnetic fields. (**f**) False-color plot of the exponent $n$ as a function of $B$ and $T$ with $n = d\ln(\rho_{\mathrm{xx}} - \rho_0)/d\ln T$. $Co_3In_2S_2$ clearly exhibits a NFL feature as manifested in the large orange region by $n \approx 1.5$.

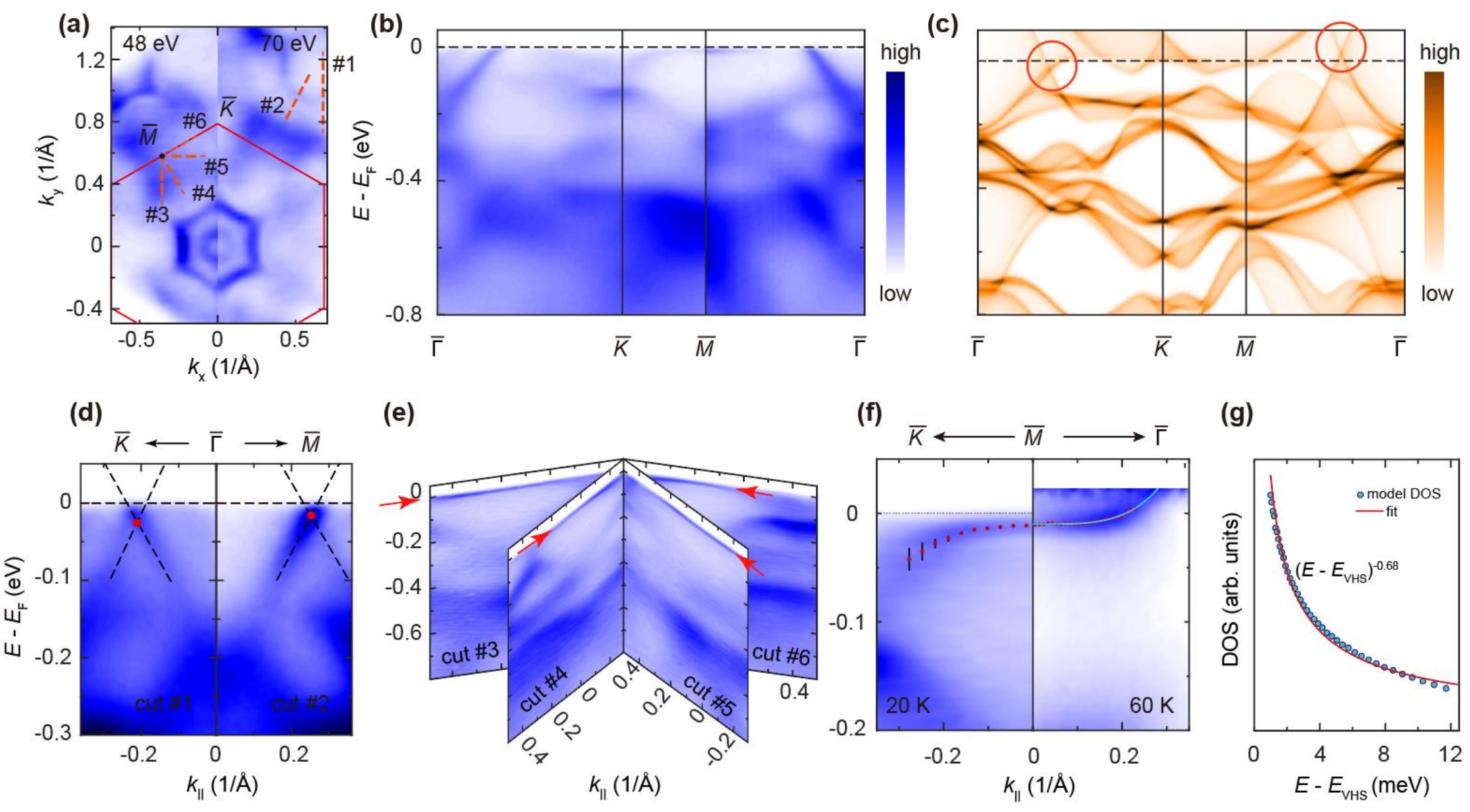

**Fig. 3 | Dirac point and high-order van Hove singularity at the Fermi level. (a)** Fermi surface measured using 48 eV (left) and 70 eV (right) photons. Data collected with linear-horizontally (LH) and linear-vertically (LV) polarized photons were merged. The red lines indicate the Brillouin zone (BZ). (**b)** Band dispersions along the high-symmetry directions ($\bar{\Gamma} - \bar{K} - \bar{M} - \bar{\Gamma}$). **(c)** Surface-projected density-functional-theory (DFT) calculations of the band structure obtained with relaxed lattice parameters, showing an overall agreement with the experiment in (**b**). **(d)** Experimental band dispersions along cut #1 and cut #2 in (**a**) showing the Dirac Fermions near $E_{\mathrm{F}}$. **(e)** Experimental band dispersions along different momentum directions near the $\bar{M}$ point (cut #3-#6 in (**a**)). Flat bands were universally observed along different momentum directions at $E_{\mathrm{F}}$ as marked by the red arrows. **(f)** Band dispersions along $\bar{K}\bar{M}$ and $\bar{M}\bar{\Gamma}$ forming a high-order van Hove singularity (HOVHS). Data along $\bar{M}\bar{\Gamma}$ were collected at 60 K and divided by Fermi-Dirac function to show the dispersion above $E_{\mathrm{F}}$. The red dots are the peak positions of the flat band extracted from Lorentzian fits to the EDCs, and the blue curve is the quartic fit $E \propto k^4$ to the electron-like branch. For the electron-like branch, the quartic fit yields an RMS deviation of ~0.42 meV, to be compared with ~1.30 meV for a quadratic fit to the same peak positions (see Supplementary Material [27]). **(g)** DOS of the HOVHS calculated from the band dispersions extracted from (f). The extracted dispersions were fitted to polynomial

functions and the DOS was evaluated from its definition with $d = 2$, neglecting the weak $k_z$ dispersion near the band bottom. The red curve is the fit of the DOS to $(E - E_{\mathrm{VHS}})^{-\beta}$ with $\beta = 0.68 \pm 0.02$.

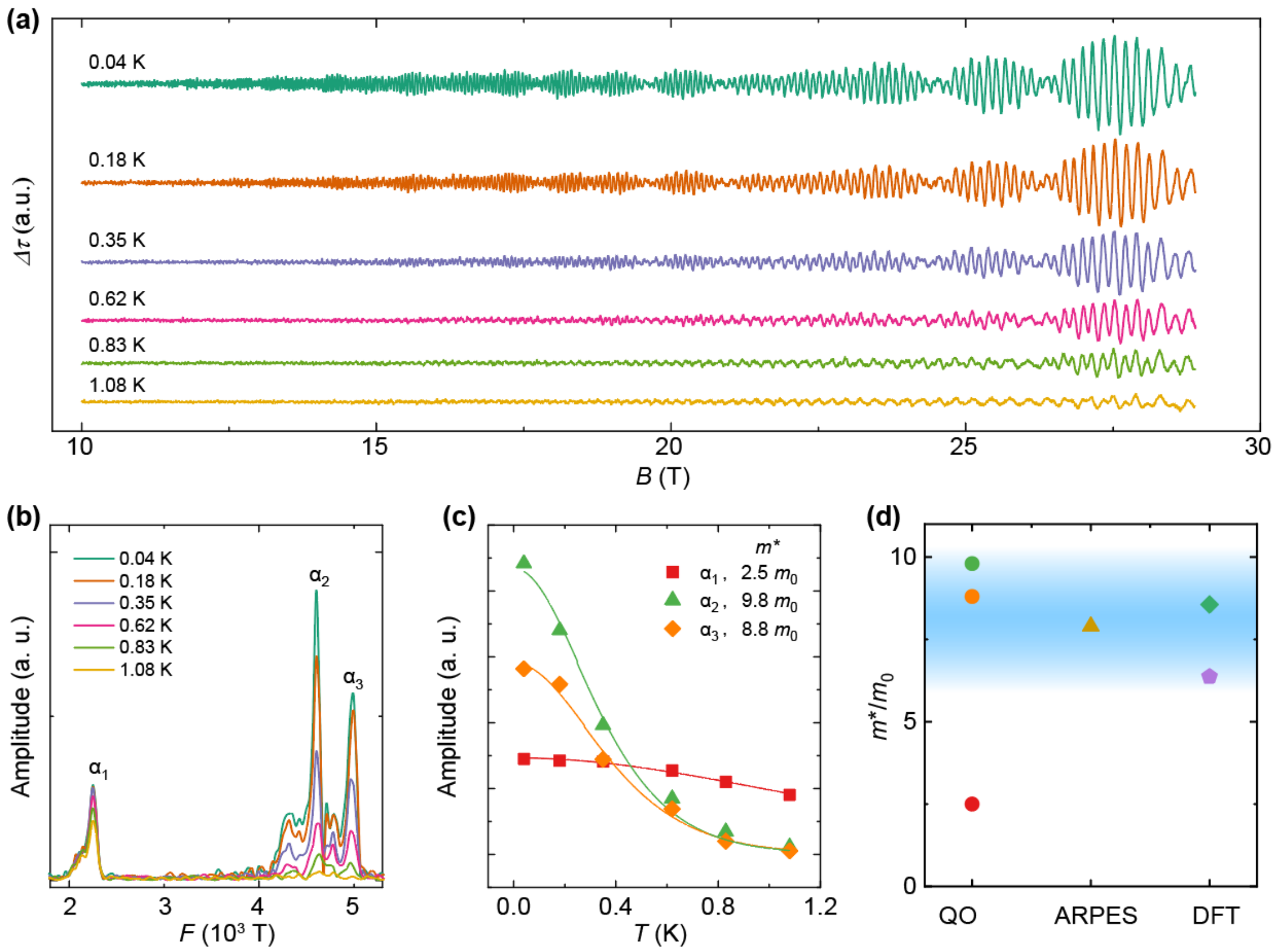

**Fig. 4 | de Haas–van Alphen oscillations of magnetic torque**. (a) Magnetic-field dependence of the oscillatory torque signal after subtracting the non-oscillatory background using polynomial fitting. (b) FFT spectrum of the oscillatory component in the high-field range of 16-29 T, revealing multiple frequencies of 2248 T, 4604 T, and 4996 T, labeled as $\alpha_1$, $\alpha_2$, $\alpha_3$. (c) Temperature dependence of the FFT amplitude for each frequency, fitted with the Lifshitz–Kosevich formula to extract the cyclotron effective masses. (d) Comparison of the effective masses obtained from quantum oscillation (QO), ARPES measurements, and DFT calculations. The ARPES and DFT values are estimated from the HOVHS band at $E_F$; the two DFT values correspond to two different $k_z$ points of the HOVHS band at $E_F$ [27]. All three independent probes consistently yield an effective mass of $m^* \approx$ 6-10 $m_0$**,** confirming that the HOVHS-derived quasiparticles dominate the thermodynamic response and exhibit heavy-electron behavior.

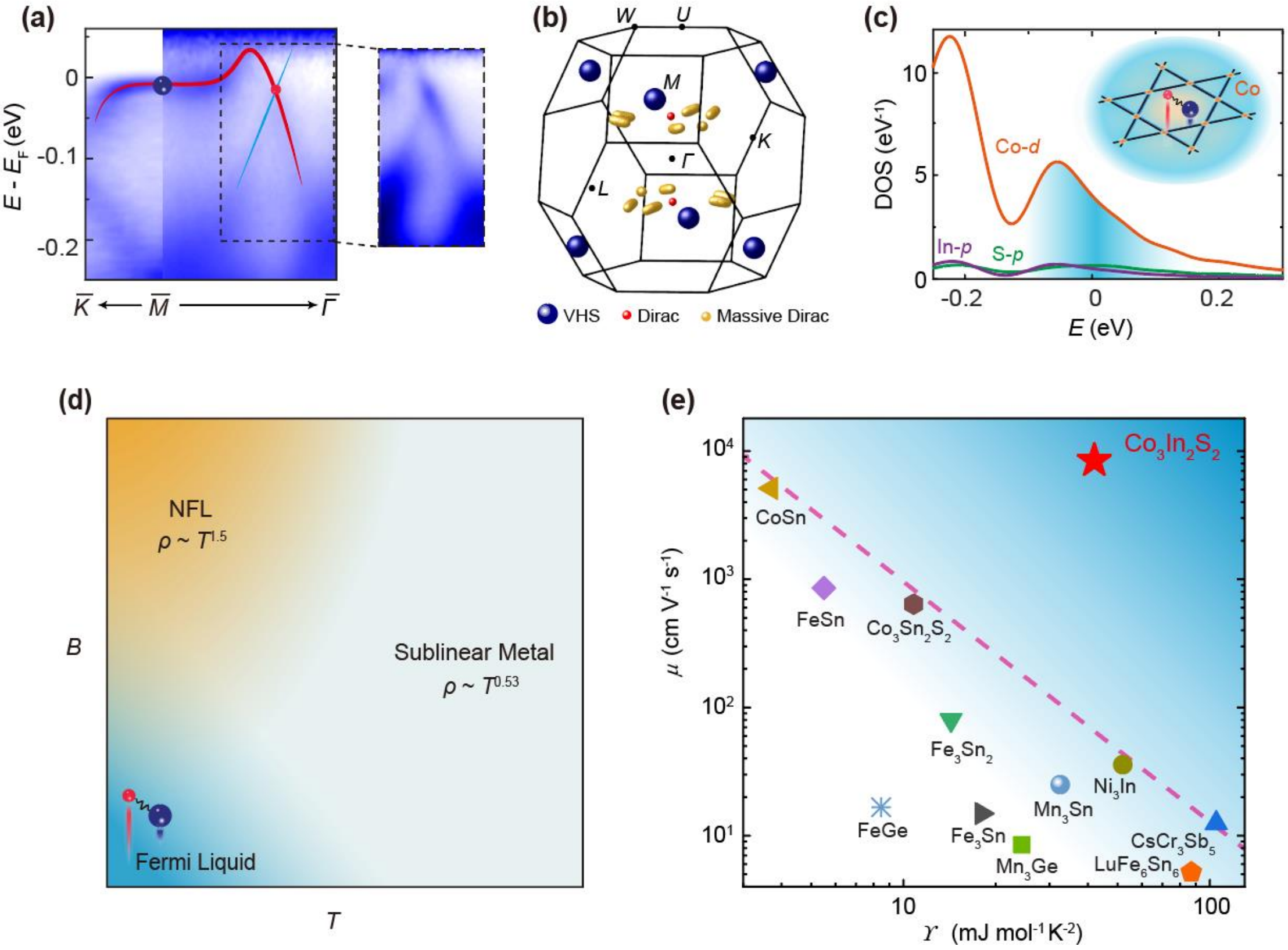

**Fig. 5 | Duality of fast and heavy carriers. (a)** Experimental band structure showing the entangled Dirac fermion and HOVHS. The Dirac fermion is better visualized in the second BZ as shown in the dashed rectangle. (**b)** Calculated Dirac points and VHSs distributed in the Brillouin zone. Massive Dirac denotes the anti-crossing nodal bands with gaps smaller than 10 meV. (**c)** Calculated DOS of Co *d*, In *p*, and S *p* orbitals. It is evident that Co-3*d* electrons dominate the total DOS around the Fermi level. The inset schematically demonstrates the fascinating physics in $Co_3In_2S_2$: a kagome lattice accommodates heavy electrons from HOVHS interacting with Dirac fermions, giving rise to anomalous transport behaviors. (**d)** Schematic *T*-*B* phase diagram summarizing the transport properties of $Co_3In_2S_2$. At low temperatures and low magnetic fields, both the high mobility of the Dirac fermions and large effective mass of heavy electrons from HOVHS were evident in the experiments. With increasing temperature and magnetic fields, a sublinear transport behavior emerges above about 100 K. At high magnetic fields and low temperatures, a typical NFL behavior was observed. **(e)** Comparison of mobility $\mu$ and $\gamma$ of different kagome materials highlighting the uniqueness of $Co_3In_2S_2$ featuring the duality with both high mobility and large effective

mass. The pink dashed line is a guide to the eye, indicating a general trend in regular kagome compounds: materials with a large (small) $\gamma$ usually exhibits a small (large) $\mu$.